\documentclass[11pt]{article}
\usepackage{tikz}
\usepackage{latexsym,amsthm,amsmath,amssymb,url}
\usepackage[colorlinks=true,citecolor=blue]{hyperref}
\usepackage[nameinlink]{cleveref}
\usepackage{fullpage}

\newcommand{\prf}{{\rm prf}}
\newcommand{\cost}{{\rm cost}_{\lambda}}
\newcommand{\FPT}{{\sf FPT}}
\newcommand{\NP}{{\sf NP}}
\newcommand{\Po}{{\sf P}}
\newcommand{\XP}{{\sf XP}}
\newcommand{\Wo}{{\sf W}}
\newcommand{\bO}{\mathcal{O}}

\newtheorem{openq}{Open Problem}
\newtheorem{theorem}{Theorem}

\newtheorem{lemma}{Lemma}

\title{A Survey on Graph Problems\\ Parameterized Above and Below Guaranteed Values}
\author{Gregory Gutin\thanks{Royal Holloway, University of London, Department of Computer Science, Egham, United Kingdom, \texttt{gutin@cs.rhul.ac.uk}}
  \and Matthias Mnich\thanks{Hamburg University of Technology, Institute for Algorithms and Complexity, Hamburg, Germany, \texttt{matthias.mnich@tuhh.de}}}
\date{}

\begin{document}

\maketitle

\begin{abstract}
  We survey the field of algorithms and complexity for graph problems parameterized above or below guaranteed values.
  Those problems seek, for a given graph $G$, a solution whose value is at least $g(G) + k$ or at most $g(G) - k$, where $g(G)$ is a guarantee on the value that \emph{any} solution on $G$ takes.
  The goal is to design algorithms which find such solution in time whose complexity in $k$ is decoupled from that in the guarantee, or to rule out the existence of such algorithms by means of intractability results.
  We discuss a large number of algorithms and intractability results, and complement them by several open problems.
\end{abstract}

\noindent
{\bf Keywords:} parameterized problems; graphs; paramerization above guarantee; paramerization below guarantee; fixed parameter tractable

\section{Introduction}
\label{sec:intro}
A classical problem in algorithmic graph theory is the {\sc Max Cut} problem.
A {\em cut} of a graph $G=(V,E)$ is the set  $(U,V\setminus U)$ of edges with exactly one endpoint in some proper subset $U$ of $V$.
In the {\sc Max Cut} problem, given a graph $G$, the aim is to find a cut of maximum size, where the \emph{size} of a cut it its number of edges).
{\sc Max Cut} belongs to Karp's famous list of 21 $\mathsf{NP}$-hard problems~\cite{Karp1972}.
As such, it is acceptable to design algorithm for {\sc Max Cut} which spend superpolynomial time to find optimal solutions.
Indeed, the best known algorithms for {\sc Max Cut} in terms of the number $n = |V|$ of nodes and $m = |E|$ of edges take times $2^{\bO(n)}$ and $2^{\bO(m)}$.
In fact, under the fundamental Exponential Time Hypothesis~\cite{ImpagliazzoPZ2001}, the problem cannot be solved in time which is subexponential in $n$.
This intractability result motivates the design of algorithms which confine the exponential-time part of their run time to some parameters $p(G)$ which are always at most as large as $n$ on any graph $G$.
We can parameterize\footnote{We give an introduction to parameterized algorithms and complexity in \Cref{sec:tn}.} {\sc Max Cut} in different ways. 
In the standard parameterization of {\sc Max Cut}, denoted by \mbox{$k$-{\sc Max Cut}}, we are to decide whether $G$ has a cut of size at least $k$, where~$k$ is the parameter.
However, Mahajan and Raman  \cite{MahajanR1997,MahajanR1999} observed that the standard parameterization of {\sc Max Cut} is not in the spirit of parameterized complexity.
Indeed, it is well-known that~$G$ always has a cut of size at least $\lceil m/2\rceil$.
Thus, for $k\le \lceil m/2\rceil$ every instance of $k$-{\sc MaxCut} is a ``yes''-instance, and therefore only for $k>\lceil m/2\rceil$ the problem is of any interest.
However, then the parameter $k$ is quite large and for so large parameters ``fixed-parameter tractable algorithms are infeasible'' \cite{MahajanR1997,MahajanR1999}. 

Also, it is easy to see that $k$-{\sc Max Cut} has a kernel with a linear number of edges.
Indeed, consider an instance $G$ of $k$-{\sc MaxCut}.
As we mentioned above, if $k\le \lceil m/2\rceil$ then $G$ is a ``yes''-instance; otherwise, we have $k>\lceil m/2\rceil$ and $m\le 2k$.
Such a kernel should be viewed as {\em large} rather than {\em small}, as the bound~$2k$ might suggest at the first glance (since $k$ is large). 

Thus, Mahajan and Raman~\cite{MahajanR1997,MahajanR1999} noted that a ``more meaningful question in a parameterized setting is to ask whether $\dots$ $\lceil m/2 \rceil + k$ edges can be placed in a cut,'' where $k$ is the parameter.
Note that this parameterization is one \emph{above guarantee}, where the guarantee equals $\lceil m/2 \rceil$.
However, Mahajun, Raman and Sikdar \cite{MahajanRS2009} observed that really meaningful questions in parameterized setting are those with guarantees being ``tight'' lower or upper bounds; here, \emph{tight} means there is an infinite family of problem instances whose optimal solution values are equal to the value of the guarantee.
For {\sc Max Cut}, a tight lower bound on the size of a maximum cut is $g(G) = m/2+(n-1)/4$, which was proved by Edwards~\cite{Edwards1975} and is usually called the Edwards-Erd\H{o}s bound due to Erd\H{o}s' contributions to the topic.
In particular, the ``right'' question to ask is whether a given connected graph $G$ on $n$ vertices and $m$ edges has a cut of size $\lceil m/2+(n-1)/4 \rceil+k$.

Using the Edwards-Erd\H{o}s bound, it is not hard to see that $k$ is still a large parameter in the parameterization $\lceil m/2 \rceil + k$, and that the parameterization
is fixed-parameter tractable and admits a polynomial-time kernel.
Thus, the guarantees/bounds that are of interest should be tight.
Most of the lower and upper bounds considered in this paper are tight and we will only point out non-tightness of bounds rather than their tightness. 

To the best of our knowledge, the works by Mahajan and Raman \cite{MahajanR1997,MahajanR1999} were the first on problems parameterized above or below tight bounds.
Since then, a large number of papers have appeared on the topic, some on graph and hypergraph problems and others on constraint satisfaction problems. 
The first survey paper on the topic was written in 2009 by Mahajan, Raman and Sikdar~\cite{MahajanRS2009}; that paper contained also a number of original results and open problems.
That publication has led to a great interest in studying problems parameterized above or below tight bounds and many interesting results on the topic were obtained, especially on constraint satisfaction problems. 
As a result, the next survey paper on the topic by Gutin and Yeo \cite{GutinY2012}.
Their survey \cite{GutinY2012} was updated later by the same authors \cite{GutinY2017}.
In the last 15 years, since the first survey, there has been a significant progress it obtaining interesting results on graph problems parameterized above or below tight bounds, and this motivates this survey paper.

Our survey briefly covers many results on the topic and states several open problems, some taken from published papers and others are new.
The paper is organized as follows.
The next section provides basics on parameterized algorithms and complexity and can be skipped by readers familiar with the area basics.
\Cref{sec:LAP} briefly describes results and an open problem on two above-guarantee (additive and multiplicative) parameterizations of the {\sc Minimum Arrangement} problem and on two additive above-guarantee parameterization of the {\sc Minimum Profile} problems.
Both problems are graph layout problems.
\Cref{sec:vc}-\ref{sec:tsp} discuss results and open problems on above/below guarantee parameterizations of the following problems, respectively: {\sc Minimum Vertex Cover}, {\sc Maximum Cut}, {\sc Maximum Bisection}, {\sc Maximum Independent Sets}, {\sc Longest Path}, {\sc Longest Cycle}, and {\sc Traveling Salesperson}.
While \Cref{sec:vc}-\ref{sec:tsp}, all but two, consider additive parameterizations, \Cref{sec:LAP}
considers multiplicative parameterizations for {\sc Minimum Arrangement} and {\sc Minimum Profile} and \Cref{sec:longest-cycle} considers multiplicative parameterizations for {\sc Longest Cycle}.
Multiplicative parameterizations for other problems are discussed in \Cref{sec:mult}.

Note that some results on the topic are not only of theoretical interest. Indeed, we will briefly discuss the paper of Ferizovic et al. \cite{FerizovicHLMSS2020} where the authors introduced practical computing reduction rules for {\sc Max Cut} mainly based on such rules used for solving {\sc Max Cut} parameterized above the Edwards-Erd\H{o}s bound.

\section{Basics on Parameterized Algorithms and Complexity} 
\label{sec:tn}
We consider \emph{parameterized problems} $\Pi$ with bivariate inputs $(I,k)$, where $I$ is the \emph{problem instance} and $k\in\mathbb N$ is the \emph{parameter}.
Throughout, we assume that~$k$ is polynomially bounded in $|I|$, the \emph{size} of $I$.
As our focus is on graph problems, we have that $I$ is a (directed or undirected) graph $G = (V,E)$; throughout, we set $n = |V|$ and $m = |E|$.
Instead of directed graphs, we simply talk about digraphs.

We say that $\Pi$ is \emph{fixed-parameter tractable} if membership of $(I,k)$ in $\Pi$ can be decided by an algorithm of run time $f(k)|I|^{\bO(1)}$, where $f(k)$ is an arbitrary function of the parameter $k$ only.
Such an algorithm is called a \emph{fixed-parameter algorithm}, and problem $\Pi$ is said to belong to complexity class $\FPT$.
In this paper, we will sometimes use a shortcut $\bO^*(f(k))$ for $f(k)|I|^{\bO(1)}$, i.e., $\bO^*$ hides not only constant factors, but also polynomial factor which are polynomial in $n$.

If the non-parameterized version of $\Pi$ (where $k$ is just a non-distinguished part of the input) is \NP-hard, then the function $f(k)$ must be superpolynomial
provided \Po$\neq$\NP.
Often, $f(k)$ is ``moderately exponential,'' which can make the problem practically tractable for small values of~$k$.
Thus, it is important to parameterize a problem in such a way that the instances with small values of $k$ are of real interest.

When the run time is relaxed by the much more generous $|I|^{\bO(f(k))}$, we obtain the class $\XP$ of problems which are polynomial-time solvable for any \emph{fixed} value of $k$.
Hence, $\FPT \subseteq \XP$.

Let $\Pi$ and $\Pi'$ be parameterized problems with parameters $k$ and $k'$, respectively.
An \emph{\FPT-reduction $R$ from~$\Pi$ to $\Pi'$} is a many-to-one transformation from~$\Pi$ to $\Pi'$, such that (i) $(I,k)\in \Pi$ if and only if $(I',k')\in \Pi'$ with $k'\le g(k)$ for a fixed computable function $g$, and (ii) $R$ is of complexity $\bO^*(f(k))$.
To show that a problem $\Pi'$ is unlikely is to be fixed-parameter tractable, one commonly provides an \FPT-reduction from a problem $\Pi$ that is $\mathsf{W}[1]$-hard; then $\Pi'$ is also $\mathsf{W}[1]$-hard.
Under the fundamental hypothesis $\mathsf{FPT}\not=\mathsf{W}[1]$, this means that problem $\Pi'$ is not fixed-parameter tractable.
A typical $\mathsf{W}[1]$-hard problem~$\Pi$ that one reduces from to show $\mathsf{W}[1]$-hardness for graph problems is {\sc Clique} paramterized by the size $k$ of the clique which one seeks in a given input graph.

We further say that $\Pi$ is in \emph{para-\NP} if membership of $(I,k)$ in $\Pi$ can be decided in nondeterministic time $f(k)|I|^{\bO(1)}$, where $f(k)$ is an arbitrary function of the parameter $k$ only; here, nondeterministic time means that we can use nondeterministic Turing machine.
A parameterized problem $\Pi'$ is {\em para-\NP-complete} if it is in para-\NP, and for any parameterized problem $\Pi$ in para-\NP~there is an \FPT-reduction from $\Pi$ to $\Pi'$.

Given a pair $\Pi,\Pi'$ of parameterized problems, a \emph{bikernelization from $\Pi$ to $\Pi'$} is a polynomial-time algorithm that maps an instance $(I,k)$ to an instance $(I',k')$ (the \emph{bikernel}) such that\linebreak \mbox{(i)~$(I,k)\in \Pi$} if and only if $(I',k')\in \Pi'$, (ii)~ $k'\leq f(k)$, and (iii)~$|I'|\leq g(k)$ for some functions~$f$ and $g$.
The function $g(k)$ is called the {\em size} of the bikernel.
A {\em kernelization} of a parameterized problem $\Pi$ is simply a bikernelization from $\Pi$ to itself, in which case $(I',k')$ is called a {\em kernel}.
The term bikernel was coined by Alon et al. \cite{AlonGKSY2011}; Bodlaender et al.~\cite{BodlaenderTY2011} refer to it as a generalized kernel.

It is well-known that a parameterized problem $\Pi$ is fixed-parameter tractable if and only if it is decidable and admits a kernelization \cite{CyganFKLMPPS2015,DowneyF2013}.
This equivalence can be easily extended as follows: A decidable parameterized problem $\Pi$ is fixed-parameter tractable if and only if it admits a bikernelization from itself to a decidable parameterized problem $\Pi'$ \cite{AlonGKSY2011}.

Due to applications, low-degree polynomial size kernels are of main interest.
Unfortunately, many fixed-parameter tractable problems do not have kernels of polynomial size, unless ${\sf NP} \subseteq {\sf coNP/poly}$~\cite{CyganFKLMPPS2015,DowneyF2013}.
It is well-known \cite{AlonGKSY2011} that the existence of a polynomial bikernel for a problem implies the existence of a polynomial kernel.


Many lower bound results for parameterized complexity have been proved under the assumption of the Exponential Time Hypothesis~\cite{ImpagliazzoPZ2001}: the hypothesis states that 3-SAT cannot be solved in $\bO(2^{\delta n})$ time for some $\delta > 0,$ where $n$ is the number of variables in the CNF formula of 3-SAT.

For further background and terminology on parameterized complexity, we refer the reader to the monographs by Cygan et al.~\cite{CyganFKLMPPS2015} and Downey and Fellows~\cite{DowneyF2013}.

\section{Graph Layout Problems}
\label{sec:LAP}
There are many graph layout problems of interest in applications. 
A survey by Serna and Thilikos \cite{SernaT2005} on parameterized complexity of graph layout problems lists ten such problems. 
In this section we consider above guarantee parameterizations of two graph layout problems: {\sc Minimum Arrangement} and {\sc Minimum Profile}.
The following is the key notion for graph layout problems.
An \emph{ordering} of a graph $G=(V,E)$ is a bijection $\alpha:V\rightarrow \{1,\hdots,n\}$.

\subsection{Minimum Arrangement}\label{sec:la}
Given an ordering $\alpha$ of $G$, the \emph{length} of an edge $\{u,v\}\in E$ is defined as
\begin{equation*}
   \lambda_\alpha(uv)=|\alpha(u)-\alpha(v)| \enspace .
\end{equation*}
The \emph{cost} $\cost(\alpha,G)$ of an ordering $\alpha$ is the sum of lengths of all edges of $G$ relative to $\alpha$, i.e.,
\begin{equation*}
  \cost(\alpha,G)=\sum_{e\in E} \lambda_\alpha(e) \enspace .
\end{equation*}
The \emph{minimum ordering cost} $\cost(G)$ is the minimum of $\cost(\alpha,G)$ over all linear arrangements~$\alpha$ of $G$.

In the {\sc Minimum Arrangement} problem, given a graph $G=(V,E)$ and an integer $k$, one has to decide whether $\cost(G)\le k$.
This problem is known to be {\sf NP}-complete \cite{GareyJ1979}. 
Note that for every ordering $\alpha$ it holds $\cost(\alpha,G)\ge m$.
This bound is attained by the family $(P_n)$ of paths of length $n$, for all $n\in\mathbb N$.
Hence, $g(G) = m$ is a guarantee.
Thus, {\sc Minimum Arrangment} parameterized by the cost $k$ an optimal ordering is trivially fixed-parameter tractable.
Thus, Fernau~\cite{Fernau2005} asked whether the natural above guarantee parameterization of {\sc Minimum Arrangement} by $k = \cost(G) - m$ is fixed-parameter tractable.
Gutin et al. \cite{GutinRSY2007} proved the following:

\begin{theorem}
\label{th:MAfpt}
  There is an algorithm that in time $\bO(n+m+5.88^k)$ decides whether $\cost(G)\le m+k$.
\end{theorem}

To prove \Cref{th:MAfpt}, Gutin et al. \cite{GutinRSY2007} showed a number of lemmas linking the minimum cost of a ordering with structure of $G$.
Here are two such lemmas, which may be of independent interest. 

\begin{lemma}
  Let $G$ be a connected bridgeless graph. Then $\cost(G)\ge m+\frac{n-1}{2}$.
\end{lemma}

A bridge in a graph $G$ is {\em $k$-separating} if each of the connectivity components of $G-e$ has more thank $k$ vertices. 
\begin{lemma}
  Let $G$ be a connected graph with $\cost(G)\le m+k.$ Then either $G$ has a $k$-separating bridge or $n\le 4k+1.$
\end{lemma}

The run time in \Cref{th:MAfpt} seems far from best possible; we thus would like to state following:
\begin{openq}
  Design a faster fixed-parameter algorithm for {\sc Minimum Arrangement} parameterized by $k = \cost(G) - m$.
\end{openq}

Serna and Thilikos \cite{SernaT2005} asked about the complexity of the stronger parameterizations: decide whether $\cost(G)\le kn$ and $\cost(G)\le km$, where $k$ is the parameter.
Gutin et al. \cite{GutinRSY2007} proved the following:
\begin{theorem}
\label{thm:costNP}
  For every fixed $k\ge 2$, it is \NP-hard to decide whether $\cost(G)\le kn$. The same result holds for $\cost(G)\le km.$ 
\end{theorem}

\subsection{Minimum Profile}
The \emph{profile} of an ordering $\alpha$ of $G$ is $\prf_{\alpha}(G)=\sum_{v\in V}(\alpha(v)-\min\{\alpha(u):\ u\in N[v]\})$; here $N[v]$ denotes the closed neighborhood of~$v$.
The \emph{profile} $\prf(G)$ of~$G$ is the minimum of $\prf_{\alpha}(G)$ over all orderings $\alpha$ of $G$.
It is well-known~\cite{Billionnet1986} that $\prf(G)$ equals the minimum number of edges in an interval graph $H$ that contains~$G$ as a subgraph.
The problem of computing the minimum number of edges in $H$ is well-known to be \NP-hard~\cite{GareyJ1979}.
Thus, the problem of computing $\prf(G)$ is also \NP-hard. 
  
Moreover, $\prf(G)\ge m$ for any graph $G$, and the bound is tight as equality holds whenever $G$ is an interval graph.
Thus, the parameterized problem of deciding whether $\prf(G)\le k$, where $k$ is the parameter, is trivially fixed-parameter tractable. 
Therefore, it is natural to consider the problem of deciding $\prf(G)\le m+k$, where $k$ is the parameter.
Kaplan et al. \cite{KaplanST1999} were the first to ask what the parameterized complexity of this problem is.
That question was answered by Villanger et al.~\cite{VillangerHPT2009}, who proved the following:
\begin{theorem}
  We can decide whether $\prf(G)\le m+k$ in time $\bO(k^{2k}n^3m).$
\end{theorem}
The algorithm by Villanger et al. \cite{VillangerHPT2009} was somewhat surprising at the time of its publication as it used the method of bounded search trees, which were thought be of no interest for designing algorithms for that problem.
It performs a bounded search among the possible ways of adding edges to a graph to obtain an interval graph, in combination with a greedy algorithm when graphs of a certain structure are reached by the search.
 
Earlier, Gutin et al. \cite{GutinSY2008} proved fixed-parameter tractability of the following weaker parameterization: Given a connected graph $G$, decide whether $\prf(G)\le n-1+k$.
In that parameterization, the bound $\prf(G)\ge n-1$ for a connected graph~$G$ is used instead of $\prf(G)\ge m$ since $m\ge n-1$ (as $G$ is connected).
Note that the bound $\prf(G)\ge n-1$ is tight, as the equality holds for paths~$P_n$.
Gutin et al. \cite{GutinSY2008} showed the following:
\begin{theorem}
  For a connected graph $G$, we can decide whether $\prf(G)\le n-1+k$ in time\linebreak $\bO(n^2+(12k+6)!)$.
\end{theorem}

Serna and Thilikos \cite{SernaT2005} asked about the complexity of the following parameterized problem: decide whether $\prf(G)\le kn$, where $k$ is the parameter.
Gutin et al. \cite{GutinSY2008} proved the following:
\begin{theorem}
\label{thm:prfNP}
  For every fixed $k\ge 2$, it is \NP-hard to decide whether $\prf(G)\le kn$.
\end{theorem}

\section{Minimum Vertex Cover} 
\label{sec:vc}
A {\em vertex cover} of a graph $G$ is a set of vertices $X$ such that every edge of $G$ has some vertex in~$X$.
Let ${\rm vc}(G)$ denote the minimum size of a vertex cover of $G$. 
Let ${\cal G}_B$ denote the family of graphs with maximum degree at most $B$.
Let $G\in {\cal G}_B$ and let $G$ have $m$ edges. 

Note that ${\rm vc}(G)\ge m/B$ for all graphs $G$.
Mahajan, Raman and Sikdar \cite{MahajanRS2009} observed that $m/B$ is a tight lower bound on ${\rm vc}(G)$ (indeed, consider the disjoint union of
$m/B$ stars $K_{1,B}$).
Consequently, they asked about the fixed-parameter tractability of the following variation of {\sc Vertex Cover} problem parameterized by $B + k$: Given a positive integer $B$, a non-negative integer $k$, and a graph $G\in {\cal G}_B$, decide whether ${\rm vc}(G)\le m/B+k$.
this problem is fixed-parameter tractable when parameterized by $B+k$.
Gutin et al.~\cite{GutinKLM2011} answered this question in the affirmative by showing the following:
\begin{theorem}
  If $G\in {\cal G}_B$, we can decide if ${\rm vc}(G)\le m/B+k$ in time~$\bO^*(2^{kB})$.
\end{theorem}

Let $\mu(G)$ denote the maximum size of a matching in a graph $G$.
Note that ${\rm vc}(G)\ge \mu(G)$, and~$\mu(G)$ is a tight lower bound on ${\rm vc}(G)$ (just consider a disjoint union of $K_2$'s).
It follows from a more general result of Razgon and O'Sullivan \cite{RazgonO2009} that the problem of deciding whether ${\rm vc}(G)\le \mu(G)+k$ is fixed-parameter tractable when parameterized by $k$.
(Their more general result is that, given a CNF formula $F$ with $m$ clauses such that each clause has two literals, deciding whether there is a truth assignment which satisfies at least $m-k$ clauses is fixed-parameter tractable in $k$.) 

We will return to the parameterization above $\mu(G)$ shortly, but before doing so, we will briefly discuss the following related problem.
Note that $2\mu(G)$ is an upper bound on ${\rm vc}(G)$, whose tightness follows from a disjoint union of triangles.
Thus, one can consider the problem of deciding, for a graph $G$ and an integer $k\in\mathbb N_0$, whether ${\rm cv}(G)\le 2\mu(G)-k$.
Gutin et al. \cite{GutinKLM2011} proved that this problem is {\sf W}[1]-hard when parameterized by $k$.

Lokshtanov et al. \cite{LokshtanovNRRS14} considered the following linear programming relaxation of the well-known integer programming formulation of the problem of computing ${\rm vc}(G)$: 
\begin{eqnarray*}
\min & \sum_{v\in V(G)} x_v &\\
\mbox{subject to} & x_u+x_v\ge 1 & \forall uv\in E(G),\\
& 0\le x_v\le 1 & \forall v\in V(G).
\end{eqnarray*}
Let ${\rm vc}^*(G)$ denote the minimum value of the objective function of the linear relaxation above.
Clearly, ${\rm vc}(G)\ge {\rm vc}^*(G)$.
By the weak duality of linear programs, it follows that ${\rm vc}^*(G)\ge \mu(G)$ (for details, see e.g. Exercise 2.24 in the book by Cygan et al.~\cite{CyganFKLMPPS2015}).
Since $\mu(G)$ is a tight lower bound on ${\rm vc}(G)$, we have that ${\rm vc}^*(G)$ is a tight lower bound on ${\rm vc}(G).$
Thus, it is natural to ask whether deciding ${\rm vc}(G)\le {\rm vc}^*(G)+k$ is fixed-parameter tractable in $k$.
Lokshtanov et al.~\cite{LokshtanovNRRS14} answered this question in the affirmative, by proving the following:

\begin{theorem}
  There is an algorithm that decides, for any graph $G$, in time $\bO^*(2.32^k)$ whether ${\rm vc}(G)\le {\rm vc}^*(G)+k$.
\end{theorem}

Following that, Garg and Philip~\cite{GargP2016} considered an even stronger parameterization for {\sc Vertex Cover}.
Namely, they argue that $2{\rm vc}^*(G) - \mu(G)$ is a lower bound on the size of any vertex cover of a graph $G$, which they take as a guarantee $g(G)$.
They then study the parameterized complexity of deciding whether a graph~$G$ has a vertex cover of size $g(G) + k = (2{\rm vc}^*(G) - \mu(G)) + k$, and give a fixed-parameter algorithm with run time $3^k \cdot n^{\bO(1)}$ for this problem.
Their result was later complemented by a randomized kernel of size polynomial in $k$, with one-sided error, which was obtained by Kratsch~\cite{Kratsch2018}.

\section{Maximum Cut} 
\label{sec:max-cuts}
As mentioned in the introduction, the (unweighted) {\sc Max Cut} problem is to partition the vertex set of a given (connected) graph $G = (V,E)$ into two sets $U \subseteq V$ and $V \setminus U$ so as to maximize the total number of edges between those two sets.
%
As discussed in the introduction, a guarantee~$g(G)$ on the size of a cut in a connected graph $G$ is the Edwards-Erd{\H{o}}s bound~\cite{Edwards1975,Edwards1973}, which is $g(G) = m/2 + (n-1)/4$.
It was a long-standing open question whether {\sc Max Cut} is fixed-parameter tractable in $k = OPT(G) - g(G)$, where $OPT(G)$ denotes the maximum cut size in the input graph~$G$.
The problem was eventually solved by Crowston et al.~\cite{CrowstonJM2012,CrowstonJM2015}:
\begin{theorem}
  There is an algorithm which, for any connected graph $G$, decides in time $8^k\cdot \bO(n^4)$ whether~$G$ admits a cut of size at least $m/2 + (n-1)/4 + k$.
\end{theorem}
Moreover, they show the problem admits a polynomial-size kernel with $\bO(k^5)$ vertices.
The key to their result was the novel technique of one-way data reductions, which they introduced in that work and which generalizes the two-way reductions that are commonly applied for the design of fixed-parameter and kernelization algorithms.
Their result was extended by Crowston et al.~\cite{CrowstonGJM2013} to the more general {\sc Signed Max Cut} problem, where each edge is labeled either ``$+$'' or ``$-$'', and one wishes to find a cut which contains as many ``$+$'' edges and as few ``$-$'' edges as possible,
The authors also decreased the kernel size to $\bO(k^3)$ vertices~\cite{CrowstonGJM2013}.
For restricted graph classes, kernels with only $\bO(k^2)$ vertices for {\sc (Signed) Max Cut} were obtained by Faria et al.~\cite{FariaKSS2017}.
Finally, Etscheid and Mnich~\cite{EtscheidM2016} improved the kernel size on general (even signed) graphs to an optimal $\bO(k)$ vertices, and showed how to compute it in linear time $\bO(k \cdot (n+m))$.

Another guarantee $g(G)$ on the size of a cut in any (connected) graph $G$ is the number of edges in a spanning tree, that is, $g(G) = n - 1$.
Madhathil et al.~\cite{MadathilSZ2018,MadathilSZ2020} show that {\sc Max Cut} parameterized by the excess $k$ above the spanning tree lower bound admits an algorithm that runs in time $\bO(8^k)$, and hence it is fixed-parameter tractable with respect to $k$.
Furthermore, they show a polynomial kernel of size $\bO(k^5)$.
The kernelization result was significantly improved by Bliznets and Epifanov \cite{BliznetsE23} who obtained a kernel with $\bO(k)$ vertices.

Ferizovic et al.~\cite{FerizovicHLMSS2020} introduced new data reduction rules for the {\sc Max Cut} problem, which encompass nearly all previous reduction rules.
A key advantage of this generality is that their data reduction rules can be applied in a wider variety of cases—increasing their efficacy at reducing graph size.
Furthermore, they engineer efficient implementations of these reduction rules and show through extensive experiments that kernelization achieves a significant reduction on sparse graphs.
Their experiments reveal that current state-of-the-art solvers can be sped up by up to multiple orders of magnitude when combined with their data reduction rules.
On social and biological networks in particular, kernelization enabled them to solve four instances that were previously unsolved in a ten-hour time limit with state-of-the-art solvers; three of these instances where solved in less than two seconds with their kernelization.

All these works refer to unweighted {\sc Max Cut}.
For instances $G = (V,E)$ of {\sc Max Cut} with edges $e\in E$ weighted by positive integers $w(e)$, Poljak and Turz{\'i}k~\cite{PoljakT1986} proved a lower bound of $w(G)/2 + w(T_{\min})/4$, where $w(G) = \sum_{e\in E}w(e)$ and $T_{\min}$ is the minimum weight of any spanning tree of $G$.
The lower bound was recently improved by Gutin and Yeo~\cite{GutinY21} to $w(G)/2 + w(D)/4$, where $D$ is a DFS tree of $G$.
They also obtain several other polynomial-time computable lower bounds on the maximum cut size.
Yet, for none of these bounds we are aware of investigation regarding the parameterized complexity of {\sc Max Cut} parameterized above them.
\begin{openq}
  What is the parameterized complexity of {\sc Max Cut} in edge-weighted graphs parameterized above known polynomial-time computable lower bounds, such as the Poljak-Turz{\'i}k bound?
\end{openq}

\section{Maximum Bisection}
\label{sec:max-bisections}
Bisections are special kinds of bipartitions $(U,V\setminus U)$ of graphs $G = (V,E)$, where each of the two sets $U$ and $V\setminus U$ has an equal number or almost equal number of vertices; that is, $||U| - |V\setminus U|| \leq 1$.
The goal of the {\sc Max Bisection} problem is to find a bisection of maximum size, where (like in maximum cuts) the number of edges with one endpoint in $S$ and the other endpoint in $V(G)\setminus S$ is maximized.
It is not hard to see that $g(G) = \lceil m/2\rceil$ is a tight lower bound on the maximum size of a bisection of any $m$-edge graph $G$.

Gutin and Yeo~\cite{GutinY2010} initiated the study of parameterizing {\sc Max Bisection} above $g(G)$, which is to decide whether a graph $G$ has a bisection of size at least $g(G)+k$, where $k$ is the parameter.
We will call this problem {\sc Bisection above Half Edges} ({\sc BaHE}). 
Gutin and Yeo~\cite{GutinY2010} showed that {\sc BaHE} has a kernel with $\bO(k^2)$ vertices and $\bO(k^3)$ edges.
Furthermore, Gutin and Yeo~\cite{GutinY2010} designed an $\bO^*(16^k)$-time algorithm for the problem. 
Later, Mnich and Zenklusen~\cite{MnichZ2012} improved this result to a kernel with at most $16k$ vertices.
Feng et al.~\cite{FengZW2020} further improved the bound as follows.
\begin{theorem}
  {\sc BaHE} has a kernel with at most $8k$ vertices. 
\end{theorem}

The above results lead to the following:
\begin{openq}
  Find the minimum value of $c$ such that {\sc BaHE} admits a kernel with at most~$ck$ vertices. 
\end{openq}

\begin{openq}
  Design a faster fixed-parameter algorithm for {\sc BaHE} than that by Gutin and Yeo.
  Is there an $\bO^*(2^k)$-time algorithm for the problem?
\end{openq}

\section{Maximum Independent Sets}
\label{sec:independent-sets}
The {\sc Independent Set} problem seeks, for a given graph $G$, a largest set $I$ of pairwise non-adjacent vertices in $G$.
Any independent set in a graph is the complement of a vertex cover in the same graph; therefore, finding independent sets of maximum size is computationally equivalent to finding vertex covers of minimum size, and hence is $\mathsf{NP}$-hard.
The $\mathsf{NP}$-hardness of {\sc Independent Set} pertains to input graphs $G$ which are \emph{planar}, which are graphs that can be embedded in the plane such that none of their edges cross.
One of the deepest, and well-known, statements about planar graphs is the Four-Colour Theorem: it states that the vertices of any $n$-vertex planar graph $G$ can be colored with just four colours, such that any two adjacent vertices of $G$ receive distinct colours.
It took over a century of research and several falsified proofs before the first correct proof of the Four-Colour Theorem was found by Appel and Haken \cite{AppelH1976}.
The proof was later simplified by Robertson et al.~\cite{RobertsonSST1997}.
These authors also gave an $\mathcal O(n^2)$-time algorithm to actually find such a 4-colouring.
Any colour class in such 4-colouring forms an independent set of $G$; in particular, the largest colour class forms an independent set of size at least $\lceil n/4\rceil$.
To date, the 4-colouring algorithm is the only known polynomial-time algorithm to obtain an independent set of size $\lceil n/4\rceil$ in $n$-vertex planar graphs.
This lower bound on the size of a maximum independent set is tight, as is for instance witnessed by the complete graphs $K_4$.
As of now, it is open whether the existence of a larger independent sets can be checked efficiently.
\begin{openq}
  Is there a polynomial-time algorithm to find an independent set of size at least $\lceil (n+1)/4\rceil$ in $n$-vertex planar graphs?
  Is there a fixed-parameter algorithm to find an independent set of size at least $\lceil (n+k)/4\rceil$ in $n$-vertex planar graphs?
\end{openq}
The {\sc Independent Set} problem remains $\mathsf{NP}$-hard in planar graphs with maximum degree at most 3, for which the lower bound of $\lceil n/4 \rceil$ is still tight.
Mnich~\cite{Mnich2016} answered the open question for planar graphs with maximum degree at most 3, by giving a fixed-parameter algorithm with run time $2^{\bO(k)}\cdot \bO(n)$.

The corresponding question for \emph{triangle-free} planar graphs was addressed by Dvorak and Mnich.
In triangle-free planar graphs, the {\sc Independent Set} problem remains $\mathsf{NP}$-hard.
A classical result by Gr{\"o}tzsch~\cite{Grotzsch1959} states that any $n$-vertex triangle-free planar graphs admits a 3-colouring, which implies the existence of an independent set of size at least $\lceil n/3 \rceil$.
Dvo\v{r}\'{a}k and Mnich~\cite{DvorakM2014} proved the following:

\begin{theorem}\label{thm:DM}
  There is an algorithm which, for any triangle-free planar graph~$G$, decides in time $2^{\bO(\sqrt{k})}\cdot \bO(n)$ whether $G$ has an independent set of size at least $\lceil (n+k)/3 \rceil$.
\end{theorem}

The algorithm of \Cref{thm:DM} relies on a reduction to graphs whose treewidth is bounded by~$\bO(k)$, on which the {\sc Independent Set} problem can eventually be solved in $2^{\bO(\sqrt{k})}\cdot \bO(n)$ time.

\section{Longest Path}
\label{sec:longest-path}
The {\sc Longest Path} problem is a classical optimization problem in graphs: given a graph $G$, one seeks a path of length at least $k$ in $G$, or concludes that no such path exists.
In their fundamental work on {\sf LogNP}-completeness, Papadimitriou and Yannakakis~\cite{PapadimitriouY1996} asked whether one can decide in polynomial time whether any given $n$-vertex graph admits a path of length $\log n$.
The question was positively resolved by Alon et al.~\cite{AlonYZ1995} when they introduced the color-coding method.
Through this method, one can find paths of length $k$ in time $2^{O(k)}\cdot n^{O(1)}$, and hence paths of length $k = \log n$ in polynomial time.
Since then, various improved algorithms have been suggested for the {\sc Longest Path} problem when parameterized by the length $k$ of the path.

Some of these algorithms also allow to specify the end points of the path, that is, one seeks a path of length (exactly or at least) $k$ that starts in some vertex $s$ and ends in some vertex $t$.
A natural lower bound on the length $k$ of such an $(s,t)$-path is the length $\ell(G,s,t)$ of a shortest $(s,t)$-path.
This lower bound is tight, as is witnessed for instance by graphs $G$ which are paths and setting $s$ and $t$ to be the endpoints of this path.
Bezekova et al.~\cite{BezakovaCDF2019} considered the following parameterization of {\sc Longest Path} above the guarantee $\ell(G,s,t)$ called {\sc Longest Detour}: For given vertices $s$ and $t$ of a graph $G$ and non-negative integer $k$, decide whether $G$ has an $(s,t)$-path that is at least $k$ longer than a shortest $(s,t)$-path.
Using insights into structural graph theory, Bezekova et al.~\cite{BezakovaCDF2019} proved that {\sc Longest Detour} is fixed-parameter tractable on undirected graphs, and can be solved in time $2^{O(k)} \cdot n^{O(1)}$.
The run time was recently improved by Fomin et al.~\cite{FominGLSSS23}, who proved the following:
\begin{theorem}
  {\sc Longest Detour} on undirected graphs can be solved in time $\bO^*(45.5^k)$ by a deterministic algorithm and in time $\bO^*(10.8^k)$ by a bounded-error randomized algorithm. 
\end{theorem}

For digraphs, the parameterized complexity of  {\sc Longest Detour} remains unknown.
\begin{openq}[\cite{BezakovaCDF2019}]
  Is {\sc Longest Detour} fixed-parameter tractable on digraphs?
\end{openq}

For a fixed integer $p$, the {\sc $p$-Disjoint Paths} problem asks for a given digraph $D$ and~$p$ pairs $(s_i,t_i)$ of its vertices whether $D$ has $p$ pairwise internally vertex-disjoint paths $Q_1,\dots ,Q_p$ such that~$Q_i$ starts at $s_i$ and terminates at $t_i$, for $i = 1,\dots,p$.
A digraph $H$ is {\em semicomplete} if there is an arc between every pair of distinct vertices of $H$ (semicomplete digraphs generalize tournaments where there is exactly one arc between every pair of distinct vertices).
Schrijver \cite{Schrijver1994} proved that  {\sc $p$-Disjoint Paths} on planar digraphs can be solved in time $n^{\bO(p)}$. Chudnovsky et al. \cite{ChudnovskySS19} proved that if we are given a partition of the vertex set of a digraph $D$ into $c$ subsets such that each subset induces a semicomplete subdigraph of $D$, then we can solve {\sc $p$-Disjoint Paths} on $D$ in time~$n^{\bO((cp)^5)}$.  

Fomin et al.~\cite{FominGLSSS23} proved that {\sc Longest Detour} is fixed-parameter tractable for every class of digraphs where {\sc 3-Disjoint Paths} can be solved in polynomial time.
This and Schrijver's result \cite{Schrijver1994} imply that for \emph{planar} digraphs there is an algorithm for solving {\sc Longest Detour} in time $2^{\bO(k)} \cdot n^{\bO(1)}$ \cite{FominGLSSS23}.
Using the result of Chudnovsky et al. \cite{ChudnovskySS19} we can get a fixed-parameter algorithm for {\sc Longest Detour} on digraphs whose vertex set can be partitioned into a bounded number of subsets such that each subset induces a semicomplete subdigraph. 
Improving the result of Fomin et al.~\cite{FominGLSSS23}, Jacob et al. \cite{JacobWZ2023a} proved that {\sc Longest Detour} is fixed-parameter tractable for every class of digraphs where {\sc 2-Disjoint Path} can be solved in polynomial time.



If one is looking for a long path (without fixed end-vertices) above a tight lower bound, then the following variation of  {\sc Longest Detour} introduced by Fomin et al.~\cite{FominGLSSS23} is of interest.
In {\sc Longest Path above Diameter}, given a graph $G$ and a nonnegative integer $k$, decide whether $G$ has a path with length at least ${\rm diam}(G)+k$, where ${\rm diam}(G)$ is the diameter of $G$ i.e., the maximum over $\ell(G,s,t)$ for all vertices $s$ and~$t$ of $G$.
Note that we cannot reduce {\sc Longest Path above Diameter} to {\sc Longest Detour} due to the following remark.
Let~$s$ and $t$ be vertices of $G$ such that the distance between them equals ${\rm diam}(G)$ and suppose that $G$ has a path $P$ of length at least ${\rm diam}(G)+k$.
Note that $P$ does not have to be an $(s,t)$-path of $G$. 

Unfortunately, {\sc Longest Path above Diameter} is {\sf NP}-complete even for $k = 1$ on undirected graphs \cite{FominGLSSS23}.
For proof, Fomin et al.~\cite{FominGLSSS23} gave an easy reduction to {\sc Longest Path above Diameter} on connected but not 2-connected graphs from {\sc Hamiltonian Path}.
However, Fomin et al.~\cite{FominGLSSS23} showed that the complexity changes if we restrict ourselves to 2-connected graphs: {\sc Longest Path above Diameter} can be solved in time $\bO^*(6.523^k)$ on undirected 2-connected graphs.

The situation is more interesting for digraphs:
\begin{theorem}[\cite{FominGLSSS23}]
\label{thm:LPaD}
  On 2-strongly-connected digraphs, {\sc Longest Path above Diameter} with $k \le 4$ can be solved in polynomial time, while for $k \ge 5$ it is {\sf NP}-complete.
\end{theorem}

A key result in proving the polynomial-time part of \Cref{thm:LPaD} is the following lemma, which is of independent interest.
\begin{lemma}
Every 2-strongly-connected digraph $G$ with ${\rm diam}(G) \ge 2^{3^{17}}$ has a path of length \mbox{${\rm diam}(G) + 4$}.
\end{lemma}

Fomin et al.~\cite{FominGLSSS23} showed that the bound ${\rm diam}(G) +4$ is best possible, as there is an infinite sequence of 2-strongly-connected digraphs $G_1, G_2, \dots$ such that for every positive integer $\ell ,$ 
${\rm diam}(G_{\ell})=8\ell +10$ and for every sufficiently large~$\ell$,  a longest path in $G_{\ell}$ has length ${\rm diam}(G_{\ell}) +4$.
This sequence of graphs is a key construction used by Fomin et al.~\cite{FominGLSSS23} to prove the {\sf NP}-completeness part of \Cref{thm:LPaD}.

We conclude this section by adding that Fomin et al. designed parameterized algorithms for computing paths above the girth of $G$ \cite{FominGLPSZ2020a} and the degeneracy of~$G$~\cite{FominGLPSZ2020b}.

\section{Longest Cycle}
\label{sec:longest-cycle}
Let $\gamma(G)$ denote the girth of a graph $G$ i.e., the minimum length of a cycle of~$G$.
Clearly, $\gamma(G)$ is a tight lower bound on the maximum length of a cycle of $G$.
Fomin et al. \cite{FominGLPSZ2021} showed that fixed-parameter tractability of deciding whether a graph $G$ has a cycle of length at least $\gamma(G)+k$ follows from that of deciding whether a graph $G$ has a cycle of length at least $k$, where $k$ is the parameter. 
Since the latter problem is well-known to be fixed-parameter tractable, Fomin et al.~\cite{FominGLPSZ2021} turned to studying the following multiplicative parameterization of {\sc Long Cycle}: Given a graph~$G$ and a non-negative integer $k$, decide whether $G$ has a cycle of length at least $k\gamma(G)$.
Due to \Cref{thm:costNP}, one may guess that the last parameterization is intractable.
However, it is tractable as proved by
Fomin et al. \cite{FominGLPSZ2021} who also studied the question of the existence of a polynomial kernel. 

\begin{theorem}[\cite{FominGLPSZ2021}]\label{thm:multcycle}
{\sc Long Cycle}  parameterized multiplicatively above girth is solvable in time $2^{\bO^{(k^2)}}n$.
However, {\sc Long Cycle} parameterized multiplicatively above girth does not admit a polynomial kernel unless {\sf NP} $\subseteq$ {\sf coNP/poly}.
\end{theorem}

To order to further strengthen the above parameterization, it is natural to replace $k$ by a slowly growing function of $\gamma(G).$ Fomin et al. \cite{FominGLPSZ2021} proved the following:
\begin{theorem}
For any fixed constant $\varepsilon > 0,$ {\sc Long Cycle} parameterized multiplicatively above $\gamma(G)^{1+\varepsilon}$ is {\sf para-NP}-hard.
\end{theorem}

The above two theorems hold also for paths rather than cycles.

What if one considers a slower growing function of $\gamma(G)$ instead of $\gamma(G)^{\varepsilon}$ as in the above theorem? 
For example, one can ask the following question. 
\begin{openq}
What is the complexity of the following problem? Given a graph $G$, decide whether $G$ has a cycle of length at least $\gamma(G)\log \gamma(G)$.
\end{openq}
Note that replacing $\log \gamma(G)$ by $\sqrt{\log \gamma(G)}$ results in a polynomial-time solvable problem by \Cref{thm:multcycle}.

Another guarantee $g(G)$ on the length of any cycle in a graph $G$ is given by the average degree~$\mathsf{ad}(G)$ (as long as it is at least 2); this result was shown by Erd{\H{o}}s and Gallai~\cite{ErdosG1959}.
Fomin et al.~\cite{FominGSS2022} addressed the parameterized complexity of {\sc Long Cycle} parameterized above the Erd{\H{o}}s-Gallai bound.
Their main result is as follows.
\begin{theorem}\label{thm:FGSS1}
  There is an $2^{\bO(k)}\cdot n^{\bO(1)}$-time algorithm which decides the existence of cycles of length $g(G)+k$ in 2-connected graphs.
\end{theorem} 
The 2-connectedness in Theorem \ref{thm:FGSS1} is necessary, as the problem is $\mathsf{NP}$-hard for $k = 1$ in general graphs.

The proof of Theorem  \ref{thm:FGSS1} uses the structural properties of dense graphs developed by Fomin et al. in \cite{FominGSS22a}. 
In 1952, Dirac proved the following classical result in graph theory: Every  $n$-vertex 2-connected graph $G$ with minimum degree $\delta\ge 2$ contains a cycle of length at least $\min\{n,2\delta\}.$
Using the structural properties of dense graphs,  Fomin et al.  \cite{FominGSS22a} proved that in time $2^{\bO(k)}\cdot n^{\bO(1)}$ one can
decide whether an $n$-vertex 2-connected graph $G$ with minimum degree $\delta\ge 2$ has a cycle of length at least $\min\{n,2\delta+k\}$.

\section{Traveling Salesperson Problem}
\label{sec:tsp}
The Travelling Salesperson Problem (TSP) is one of the most well-known and widely studied combinatorial optimization problems.
In this problem, we are given a complete graph $K_n$ with weights on its edges and we are required to find a Hamiltonian cycle in $K_n$ of minimum total weight.
In its full generality,  TSP is not only {\sf NP}-hard, but also {\sf NP}-hard to approximate to within any constant factor.
Therefore there has been much attention in developing approximation algorithms for restricted instances of TSP. 
In this section, we consider general TSP, but rather than seeking a Hamiltonian cycle of minimum weight, we seek a Hamiltonian cycle that beats the average weight of all Hamiltonian cycles by some given value.

Let $w$ be an integer edge weighting of $K_n$, i.e.\ $w: E(K_n) \rightarrow \mathbb{Z}$, and let $G$ be a subgraph of $K_n$.
We write $w(G):= \sum_{e \in E(G)} w(e)$, and define the {\it density} $d=d(w)$ of $w$ to be the average weight of an edge, i.e.\ $d:=w(K_n)/ \binom{n}{2}$.  

Vizing \cite{Vizing1973} asked whether there is a polynomial-time algorithm which, given an integer edge weighting~$w$ of $K_n$, always finds a Hamiltonian cycle $H^*$ of $K_n$ satisfying $w(H) \leq dn$.
He answered this question in the affirmative, and subsequently Rublineckii \cite{Rublineckii1973} described several other TSP heuristics satisfying this property (one of such results was first proved by E.M. Lifshits).
TSP heuristics satisfying this property are of interest, as they guarantee that the Hamiltonian cycle they construct is not longer than at least $(n-2)!$ Hamiltonian cycles if $n$ is odd and at least $(n-2)!/2$ Hamiltonian cycles if $n$ is even \cite{Rublineckii1973}.

A natural question extending Vizing's question is as follows: for each fixed~$k$, is there a polynomial-time algorithm which, given $w$, determines if there exists a Hamiltonian cycle $H^*$ satisfying $w(H^*) \leq dn - k$?
Gutin and Patel \cite{GutinP2016} answered this question affirmatively:
\begin{theorem}
\label{thm:tsp}
  There exists an algorithm which, given $(n,w,k)$ as input, where $n,k \in \mathbb{N}$ and $w: E(K_n) \rightarrow \mathbb{Z}$, determines whether there exists a Hamiltonian cycle~$H^*$ of $K_n$ with weight $w(H^*) \leq dn - k$ in time $\bO(k^3)! + \bO(k^3n) + \bO(n^7)$, and outputs such a Hamiltonian cycle if it exists.
\end{theorem}

The proof of \Cref{thm:tsp} is quite involved, and heavily uses the fact that~$K_n$ is an undirected graph.
This motivates the following problem:
\begin{openq}\label{op:tsp}
  Is there a fixed-parameter algorithm, which given a complete digraph $D$ whose every arc has an integer weight, decides whether $D$ has a Hamiltonian cycle of weight not larger that the average weight of a Hamiltonian cycle minus $k$, where $k$ is the parameter?
\end{openq}
\noindent
Gutin and Patel \cite{GutinP2016} believe that the answer to \Cref{op:tsp} is positive.

\section{Parameterizations Times Multiplicative Guarantees}
\label{sec:mult}
In the previous sections we have discussed parameterizations above/below guarantees which can be called ``additive'' parameterizations, where the parameter $k$ is taken as the additive excess over a polynomial-time computable lower bound. 
We now consider parameterizations times \emph{multiplicative} lower or upper bounds.
To the best of our knowledge, this topic was introduced by Serna and Thilikos \cite{SernaT2005}. Formally, problems of this type take as input an instance $(I, k)$ of some parameterized problem $\Pi$ with a guarantee $g(A)$, and seek a solution for $I$ which is of size at least (or at most) $k \cdot g(I)$.
Gutin et al. \cite{GutinRSY2007} and \cite{GutinSY2008} answered three open questions in \cite{SernaT2005}, see Theorems \ref{thm:costNP} and \ref{thm:prfNP}.

Fomin et al.~\cite{FominGLPSZ2021} continued such research by studying the {\sc Longest Cycle} problem parameterized multiplicatively times the girth $g(I)$ of the input graph $I = G$, see Section \ref{sec:longest-cycle}.
Apart from {\sc Longest Cycle}, Fomin et al. showed that {\sc Max Internal Spanning Tree} and {\sc Independent Set} (in graphs with girth at least 4) is fixed-parameter tractable when parameterized times the girth of the input graph, as are {\sc Vertex Cover} and {\sc Connected Vertex Cover} where one seeks solutions of size \emph{at most} $g(I)$.
Those algorithmic results are put in contrast with an intractability result, which shows that {\sc Independent Set} in general graphs is $\mathsf{W}[1]$-hard parameterized times the girth.
The authors also show that  {\sc Feedback Vertex Set} and {\sc Cycle Packing} are para-$\mathsf{NP}$-hard parameterized times the girth, though the girth is not an upper bound on the solution size~$k$ for these problems.
The algorithms heavily rely on efficient computations in bounded treewidth graphs, but differ from fixed-parameter algorithms for those problems when parameterized by (related) additive guarantees.
This fact, together with the existence of aforementioned parameterized intractability results, suggests that parameterizations times multiplicative guarantees can lead to a rich and fruitful theory of parameterized (in)tractability.

All the above results of this section hold for undirected graphs.
Fomin et al.~\cite{FominGSS2022} (and the authors of this paper in their first version) asked whether {\sc Directed Longest Cycle} is fixed-parameter tractable parameterized times the (directed) girth of the input digraph.
In answering this question, Jacob et al. \cite{JacobWZ2023b} showed that the problem is \Wo[1]-hard.



\bibliographystyle{plainurl}
\bibliography{matthias-refs,refs}

\begin{thebibliography}{10}

\bibitem{AlonGKSY2011}
Noga Alon, Gregory~Z. Gutin, Eun~Jung Kim, Stefan Szeider, and Anders Yeo.
\newblock Solving {MAX-$r$-SAT} above a tight lower bound.
\newblock {\em Algorithmica}, 61(3):638--655, 2011.

\bibitem{AlonYZ1995}
Noga Alon, Raphael Yuster, and Uri Zwick.
\newblock Color-coding.
\newblock {\em J. {ACM}}, 42(4):844--856, 1995.

\bibitem{AppelH1976}
K.~Appel and W.~Haken.
\newblock Every planar map is four colorable.
\newblock {\em Bull. Amer. Math. Soc.}, 82:711--712, 1976.

\bibitem{BezakovaCDF2019}
Ivona Bez{\'a}kov{\'a}, Radu Curticapean, Holger Dell, and Fedor~V. Fomin.
\newblock Finding detours is fixed-parameter tractable.
\newblock {\em SIAM J. Discrete Math.}, 33(4):2326--2345, 2019.

\bibitem{Billionnet1986}
A.~Billionnet.
\newblock On interval graphs and matrix profiles.
\newblock {\em RAIRO Tech. Oper.}, 20:245--256, 1986.

\bibitem{BliznetsE23}
Ivan Bliznets and Vladislav Epifanov.
\newblock Maxcut above guarantee.
\newblock In {\em Proc. MFCS 2023}, volume 272 of {\em Leibniz Int. Proc.
  Informatics}, pages 22:1--22:14, 2023.

\bibitem{BodlaenderTY2011}
Hans~L. Bodlaender, St{\'{e}}phan Thomass{\'{e}}, and Anders Yeo.
\newblock Kernel bounds for disjoint cycles and disjoint paths.
\newblock {\em Theor. Comput. Sci.}, 412(35):4570--4578, 2011.

\bibitem{ChudnovskySS19}
Maria Chudnovsky, Alex Scott, and Paul~D. Seymour.
\newblock Disjoint paths in unions of tournaments.
\newblock {\em J. Comb. Theory, Ser. {B}}, 135:238--255, 2019.

\bibitem{CrowstonGJM2013}
Robert Crowston, Gregory Gutin, Mark Jones, and Gabriele Muciaccia.
\newblock Maximum balanced subgraph problem parameterized above lower bound.
\newblock {\em Theoret. Comput. Sci.}, 513:53--–64, 2013.

\bibitem{CrowstonJM2012}
Robert Crowston, Mark Jones, and Matthias Mnich.
\newblock Max-cut parameterized above the {E}dwards-{E}rd{\H{o}}s bound.
\newblock In {\em Proc. ICALP 2012}, volume 7391 of {\em Lecture Notes in
  Comput. Sci.}, pages 242--253. 2012.

\bibitem{CrowstonJM2015}
Robert Crowston, Mark Jones, and Matthias Mnich.
\newblock Max-cut parameterized above the {E}dwards-{E}rd{\H{o}}s bound.
\newblock {\em Algorithmica}, 72(3):734--757, 2015.

\bibitem{CyganFKLMPPS2015}
Marek Cygan, Fedor~V. Fomin, Lukasz Kowalik, Daniel Lokshtanov, D{\'{a}}niel
  Marx, Marcin Pilipczuk, Michal Pilipczuk, and Saket Saurabh.
\newblock {\em Parameterized Algorithms}.
\newblock Springer, 2015.

\bibitem{DowneyF2013}
Rodney~G. Downey and Michael~R. Fellows.
\newblock {\em Fundamentals of Parameterized Complexity}.
\newblock Springer, 2013.

\bibitem{DvorakM2014}
Zden\v{e}k Dvo\v{r}\'{a}k and Matthias Mnich.
\newblock Large independent sets in triangle-free planar graphs.
\newblock In {\em Proc. ESA 2014}, volume 8737 of {\em Lecture Notes Comput.
  Sci.}, pages 346--357. 2014.

\bibitem{Edwards1975}
C.~S. Edwards.
\newblock An improved lower bound for the number of edges in a largest
  bipartite subgraph.
\newblock In {\em Proc. 2nd Czechoslovak Sympos. Graph Theory}, pages 167--181,
  1975.

\bibitem{Edwards1973}
Christopher~S. Edwards.
\newblock Some extremal properties of bipartite subgraphs.
\newblock {\em Canad. J. Math.}, 25:475--485, 1973.

\bibitem{ErdosG1959}
P{\'a}l Erd{\H{o}}s and Tibor Gallai.
\newblock On maximal paths and circuits of graphs.
\newblock {\em Acta Math. Acad. Sci. Hungar.}, 10:337--–356, 1959.

\bibitem{EtscheidM2016}
Michael Etscheid and Matthias Mnich.
\newblock Linear kernels and linear-time algorithms for finding large cuts.
\newblock In {\em Proc. ISAAC 2016}, volume~64 of {\em Leibniz Int. Proc.
  Inform.}, pages Art. No. 31, 13. 2016.

\bibitem{FariaKSS2017}
Luerbio Faria, Sulamita Klein, Ignasi Sau, and Rubens Sucupira.
\newblock Improved kernels for signed max cut parameterized above lower bound
  on $(r,\ell)$-graphs.
\newblock {\em Discrete Math. \& Theoret. Comput. Sci.}, 19, 2017.

\bibitem{FengZW2020}
Qilong Feng, Senmin Zhu, and Jianxin Wang.
\newblock An improved kernel for max-bisection above tight lower bound.
\newblock {\em Theoret. Comput. Sci.}, 818:12--21, 2020.

\bibitem{FerizovicHLMSS2020}
Damir Ferizovic, Demian Hespe, Sebastian Lamm, Matthias Mnich, Christian
  Schulz, and Darren Strash.
\newblock Engineering kernelization for maximum cut.
\newblock In {\em Proc. ALENEX 2020}, pages 27--41. 2020.

\bibitem{Fernau2005}
H.~Fernau.
\newblock {\em Parameterized Algorithmics: A graph-theoretic approach}.
\newblock Habilitation thesis, University of T{\"u}bingen, 2005.

\bibitem{FominGLSSS23}
Fedor~V. Fomin, Petr~A. Golovach, William Lochet, Danil Sagunov, Saket Saurabh,
  and Kirill Simonov.
\newblock Detours in directed graphs.
\newblock {\em J. Comput. Syst. Sci.}, 137:66--86, 2023.

\bibitem{FominGLPSZ2020a}
Fedor~V. Fomin, Petr~A. Golovach, Daniel Lokshtanov, Fahad Panolan, Saket
  Saurabh, and Meirav Zehavi.
\newblock Going far from degeneracy.
\newblock {\em {SIAM} J. Discret. Math.}, 34(3):1587--1601, 2020.

\bibitem{FominGLPSZ2020b}
Fedor~V. Fomin, Petr~A. Golovach, Daniel Lokshtanov, Fahad Panolan, Saket
  Saurabh, and Meirav Zehavi.
\newblock Parameterization above a multiplicative guarantee.
\newblock In {\em Proc. ITCS 2020}, volume 151 of {\em Leibniz Int. Proc.
  Informatics}, pages 39:1--39:13, 2020.

\bibitem{FominGLPSZ2021}
Fedor~V. Fomin, Petr~A. Golovach, Daniel Lokshtanov, Fahad Panolan, Saket
  Saurabh, and Meirav Zehavi.
\newblock Multiplicative parameterization above a guarantee.
\newblock {\em {ACM} Trans. Comput. Theory}, 13(3):18:1--18:16, 2021.

\bibitem{FominGSS22a}
Fedor~V. Fomin, Petr~A. Golovach, Danil Sagunov, and Kirill Simonov.
\newblock Algorithmic extensions of {D}irac's theorem.
\newblock In {\em Proc. SODA 2022}, pages 406--416, 2022.

\bibitem{FominGSS2022}
Fedor~V. Fomin, Petr~A. Golovach, Danil Sagunov, and Kirill Simonov.
\newblock Longest cycle above {E}rd{\H{o}}s-{G}allai bound.
\newblock In {\em Proc. ESA 2022}, volume 244 of {\em Leibniz Int. Proc.
  Informatics}, pages 55:1--55:15, 2022.

\bibitem{GareyJ1979}
M.~R. Garey and David~S. Johnson.
\newblock {\em Computers and Intractability: {A} Guide to the Theory of
  $\mathsf{NP}$-Completeness}.
\newblock W. H. Freeman, 1979.

\bibitem{GargP2016}
Shivam Garg and Geevarghese Philip.
\newblock Raising the bar for vertex cover: Fixed-parameter tractability above
  a higher guarantee.
\newblock In {\em Proc. SODA 2016}, pages 1152--1166, 2016.

\bibitem{Grotzsch1959}
H.~Gr{\"o}tzsch.
\newblock {Ein Dreifarbensatz f{\"u}r dreikreisfreie Netze auf der Kugel}.
\newblock {\em Wiss. Z. Martin-Luther-Univ. Halle-Wittenberg}, pages 109--120,
  1959.

\bibitem{GutinY2010}
Gregory Gutin and Anders Yeo.
\newblock Note on maximal bisection above tight lower bound.
\newblock {\em Inf. Proc. Lett.}, 110(21):966--969, 2010.

\bibitem{GutinKLM2011}
Gregory~Z. Gutin, Eun~Jung Kim, Michael Lampis, and Valia Mitsou.
\newblock Vertex cover problem parameterized above and below tight bounds.
\newblock {\em Theory Comput. Syst.}, 48(2):402--410, 2011.

\bibitem{GutinP2016}
Gregory~Z. Gutin and Viresh Patel.
\newblock Parameterized traveling salesman problem: Beating the average.
\newblock {\em {SIAM} J. Discret. Math.}, 30(1):220--238, 2016.

\bibitem{GutinRSY2007}
Gregory~Z. Gutin, Arash Rafiey, Stefan Szeider, and Anders Yeo.
\newblock The linear arrangement problem parameterized above guaranteed value.
\newblock {\em Theory Comput. Syst.}, 41(3):521--538, 2007.

\bibitem{GutinSY2008}
Gregory~Z. Gutin, Stefan Szeider, and Anders Yeo.
\newblock Fixed-parameter complexity of minimum profile problems.
\newblock {\em Algorithmica}, 52(2):133--152, 2008.

\bibitem{GutinY2012}
Gregory~Z. Gutin and Anders Yeo.
\newblock Constraint satisfaction problems parameterized above or below tight
  bounds: {A} survey.
\newblock In {\em The Multivariate Algorithmic Revolution and Beyond - Essays
  Dedicated to Michael R. Fellows on the Occasion of His 60th Birthday}, volume
  7370 of {\em Lecture Notes Comput. Sci.}, pages 257--286, 2012.

\bibitem{GutinY2017}
Gregory~Z. Gutin and Anders Yeo.
\newblock Parameterized constraint satisfaction problems: a survey.
\newblock In {\em The Constraint Satisfaction Problem: Complexity and
  Approximability}, volume~7 of {\em Dagstuhl Follow-Ups}, pages 179--203.
  2017.

\bibitem{GutinY21}
Gregory~Z. Gutin and Anders Yeo.
\newblock Lower bounds for maximum weighted cut.
\newblock {\em {SIAM} J. Discret. Math.}, 37(2):1142--1161, 2023.

\bibitem{ImpagliazzoPZ2001}
Russell Impagliazzo, Ramamohan Paturi, and Francis Zane.
\newblock Which problems have strongly exponential complexity?
\newblock {\em J. Comput. Syst. Sci.}, 63(4):512--530, 2001.

\bibitem{JacobWZ2023b}
Ashwin Jacob, Michal Wlodarczyk, and Meirav Zehavi.
\newblock Finding long directed cycles is hard even when {DFVS} is small or
  girth is large.
\newblock Technical report, 2023.
\newblock \url{https://arxiv.org/abs/2308.06145}.

\bibitem{JacobWZ2023a}
Ashwin Jacob, Michal Wlodarczyk, and Meirav Zehavi.
\newblock Long directed detours: Reduction to $2$-disjoint paths.
\newblock Technical report, 2023.
\newblock \url{https://arxiv.org/abs/2301.06105}.

\bibitem{KaplanST1999}
Haim Kaplan, Ron Shamir, and Robert~Endre Tarjan.
\newblock Tractability of parameterized completion problems on chordal,
  strongly chordal, and proper interval graphs.
\newblock {\em {SIAM} J. Comput.}, 28(5):1906--1922, 1999.

\bibitem{Karp1972}
Richard~M. Karp.
\newblock {\em Reducibility among combinatorial problems}, pages 85--103.
\newblock Plenum, New York, 1972.

\bibitem{Kratsch2018}
Stefan Kratsch.
\newblock A randomized polynomial kernelization for vertex cover with a smaller
  parameter.
\newblock {\em SIAM J. Discrete Math.}, 32(3):1806--1839, 2018.

\bibitem{LokshtanovNRRS14}
Daniel Lokshtanov, N.~S. Narayanaswamy, Venkatesh Raman, M.~S. Ramanujan, and
  Saket Saurabh.
\newblock Faster parameterized algorithms using linear programming.
\newblock {\em {ACM} Trans. Algorithms}, 11(2):15:1--15:31, 2014.

\bibitem{MadathilSZ2018}
Jayakrishnan Madathil, Saket Saurabh, and Meirav Zehavi.
\newblock Max-cut above spanning tree is fixed-parameter tractable.
\newblock In {\em Proc. CSR 2018}, volume 10846 of {\em Lecture Notes Comput.
  Sci.}, pages 244--256, 2018.

\bibitem{MadathilSZ2020}
Jayakrishnan Madathil, Saket Saurabh, and Meirav Zehavi.
\newblock Fixed-parameter tractable algorithm and polynomial kernel for max-cut
  above spanning tree.
\newblock {\em Theory Comput. Syst.}, 64(1):62--100, 2020.

\bibitem{MahajanR1997}
Meena Mahajan and Venkatesh Raman.
\newblock Parametrizing above guaranteed values: {MaxSat and MaxCut}.
\newblock {\em Electron. Colloquium Comput. Complex.}, (33), 1997.
\newblock URL:
  \url{https://eccc.weizmann.ac.il/eccc-reports/1997/TR97-033/index.html}.

\bibitem{MahajanR1999}
Meena Mahajan and Venkatesh Raman.
\newblock Parameterizing above guaranteed values: {MaxSat and MaxCut}.
\newblock {\em J. Algorithms}, 31(2):335--354, 1999.

\bibitem{MahajanRS2009}
Meena Mahajan, Venkatesh Raman, and Somnath Sikdar.
\newblock Parameterizing above or below guaranteed values.
\newblock {\em J. Comput. Syst. Sci.}, 75(2):137--153, 2009.

\bibitem{Mnich2016}
Matthias Mnich.
\newblock Large independent sets in subquartic planar graphs.
\newblock In {\em Proc. WALCOM 2016}, volume 9627 of {\em Lecture Notes Comput.
  Sci.}, pages 209--221, 2016.

\bibitem{MnichZ2012}
Matthias Mnich and Rico Zenklusen.
\newblock Bisections above tight lower bounds.
\newblock In {\em Proc. WG 2012}, volume 7551 of {\em Lecture Notes Comput.
  Sci.}, pages 184--193. 2012.

\bibitem{PapadimitriouY1996}
Christos~H. Papadimitriou and Mihalis Yannakakis.
\newblock On limited nondeterminism and the complexity of the {V-C} dimension.
\newblock {\em J. Comput. Syst. Sci.}, 53(2):161--170, 1996.

\bibitem{PoljakT1986}
Svatopluk Poljak and Daniel Turzík.
\newblock A polynomial time heuristic for certain subgraph optimization
  problems with guaranteed worst case bound.
\newblock {\em Discrete Math.}, 58(1):99--104, 1986.

\bibitem{RazgonO2009}
Igor Razgon and Barry O'Sullivan.
\newblock Almost {2-SAT} is fixed-parameter tractable.
\newblock {\em J. Comput. Syst. Sci.}, 75(8):435--450, 2009.

\bibitem{RobertsonSST1997}
Neil Robertson, Daniel~P. Sanders, Paul~D. Seymour, and Robin Thomas.
\newblock The four-colour theorem.
\newblock {\em J. Comb. Theory, Ser. {B}}, 70(1):2--44, 1997.

\bibitem{Rublineckii1973}
V.I. Rublineckii.
\newblock Estimates of the accuracy of procedures in the traveling salesman
  problem.
\newblock {\em Numer. Math. Comput. Tech.}, 4:18--23, 1973.

\bibitem{Schrijver1994}
Alexander Schrijver.
\newblock Finding $k$ disjoint paths in a directed planar graph.
\newblock {\em {SIAM} J. Comput.}, 23(4):780--788, 1994.

\bibitem{SernaT2005}
Maria~J. Serna and Dimitrios~M. Thilikos.
\newblock Parameterized complexity for graph layout problems.
\newblock {\em Bull. {EATCS}}, 86:41--65, 2005.

\bibitem{VillangerHPT2009}
Yngve Villanger, Pinar Heggernes, Christophe Paul, and Jan~Arne Telle.
\newblock Interval completion is fixed parameter tractable.
\newblock {\em SIAM J. Comput.}, 38(5):2007--2020, 2009.

\bibitem{Vizing1973}
Vadym~G. Vizing.
\newblock Values of the target functional in a priority problem that are
  majorized by the mean value.
\newblock {\em Kibernetika}, 5:76--78, 1973.
\newblock in Russian.

\end{thebibliography}

\end{document}